


\magnification=\magstephalf

\newbox\SlashedBox
\def\slashed#1{\setbox\SlashedBox=\hbox{#1}
\hbox to 0pt{\hbox to 1\wd\SlashedBox{\hfil/\hfil}\hss}#1}
\def\hboxtosizeof#1#2{\setbox\SlashedBox=\hbox{#1}
\hbox to 1\wd\SlashedBox{#2}}

\def\mathslashed#1{\setbox\SlashedBox=\hbox{$#1$}
\hbox to 0pt{\hbox to 1\wd\SlashedBox{\hfil/\hfil}\hss}#1}

\def\ifsmall{\iffalse}  
\def\titlepagefont{}  

\def\DefineTeXgraphics{%
\special{ps::[global] /TeXgraphics { } def}}  

\def\today{\ifcase\month\or January\or February\or March\or April\or May
\or June\or July\or August\or September\or October\or November\or
December\fi\space\number\day, \number\year}
\def\eatPrefix19{}
\def\Year{\expandafter\eatPrefix\the\year}
\newcount\hours \newcount\minutes
\def\monthname{\ifcase\month\or
January\or February\or March\or April\or May\or June\or July\or
August\or September\or October\or November\or December\fi}
\def\shortmonthname{\ifcase\month\or
Jan\or Feb\or Mar\or Apr\or May\or Jun\or Jul\or
Aug\or Sep\or Oct\or Nov\or Dec\fi}

\def\TimeStamp{\hours\the\time\divide\hours by60%
\minutes -\the\time\divide\minutes by60\multiply\minutes by60%
\advance\minutes by\the\time%
${\rm \shortmonthname}\cdot\if\day<10{}0\fi\the\day\cdot\the\year%
\qquad\the\hours:\if\minutes<10{}0\fi\the\minutes$}




\def\Title#1{%
\vskip 1in{\titlefont\centerline{#1}}\vskip .5in}



\newif\ifdraftmode
\newif\ifleftlabels  

\def\nolabels{\def\wrlabeL##1{}\def\eqlabeL##1{}\def\reflabeL##1{}}
\def\writelabels{\def\wrlabeL##1{\leavevmode\vadjust{\rlap{\smash%
{\line{{\escapechar=` \hfill\rlap{\sevenrm\hskip.03in\string##1}}}}}}}%
\def\eqlabeL##1{{\escapechar-1\rlap{\sevenrm\hskip.05in\string##1}}}%
\def\reflabeL##1{\noexpand\rlap{\noexpand\sevenrm[\string##1]}}}
\def\writeleftlabels{\def\wrlabeL##1{\leavevmode\vadjust{\rlap{\smash%
{\line{{\escapechar=` \hfill\rlap{\sevenrm\hskip.03in\string##1}}}}}}}%
\def\eqlabeL##1{{\escapechar-1%
\rlap{\sixrm\hskip.05in\string##1}%
\llap{\sevenrm\string##1\hskip.03in\hbox to \hsize{}}}}%
\def\reflabeL##1{\noexpand\rlap{\noexpand\sevenrm[\string##1]}}}
\nolabels

\newdimen\fullhsize
\newdimen\hstitle
\hstitle=\hsize 
\newdimen\hsbody
\hsbody=\hsize 
\newdimen\hbodyoffset
\hbodyoffset=\hoffset 
\newbox\leftpage
\def\abstract#1{#1}
\def\rotated{\special{ps: landscape}
\magnification=1000  
\baselineskip=14pt
\global\hstitle=9truein\global\hsbody=4.75truein
\global\vsize=7truein\global\voffset=-.31truein
\global\hoffset=-0.54in\global\hbodyoffset=-.54truein
\global\fullhsize=10truein
\def\DefineTeXgraphics{%
\special{ps::[global]
/TeXgraphics {currentpoint translate 0.7 0.7 scale
              -80 0.72 mul -1000 0.72 mul translate} def}}
\let\lr=L
\def\ifsmall{\iftrue}
\def\titlepagefont{\twelvepoint}
\trueseventeenpoint
\def\almostshipout##1{\if L\lr \count1=1
      \global\setbox\leftpage=##1 \global\let\lr=R
   \else \count1=2
      \shipout\vbox{\hbox to\fullhsize{\box\leftpage\hfil##1}}
      \global\let\lr=L\fi}

\output={\ifnum\count0=1 
 \shipout\vbox{\hbox to \fullhsize{\hfill\pagebody\hfill}}\advancepageno
 \else
 \almostshipout{\leftline{\vbox{\pagebody\makefootline}}}\advancepageno
 \fi}

\def\abstract##1{{\leftskip=1.5in\rightskip=1.5in ##1\par}} }

\def\linemessage#1{\immediate\write16{#1}}

\global\newcount\secno \global\secno=0
\global\newcount\appno \global\appno=0
\global\newcount\meqno \global\meqno=1
\global\newcount\subsecno \global\subsecno=0
\global\newcount\figno \global\figno=0

\newif\ifAnyCounterChanged
\let\terminator=\relax
\def\normalize#1{\ifx#1\terminator\let\next=\relax\else%
\if#1i\aftergroup i\else\if#1v\aftergroup v\else\if#1x\aftergroup x%
\else\if#1l\aftergroup l\else\if#1c\aftergroup c\else%
\if#1m\aftergroup m\else%
\if#1I\aftergroup I\else\if#1V\aftergroup V\else\if#1X\aftergroup X%
\else\if#1L\aftergroup L\else\if#1C\aftergroup C\else%
\if#1M\aftergroup M\else\aftergroup#1\fi\fi\fi\fi\fi\fi\fi\fi\fi\fi\fi\fi%
\let\next=\normalize\fi%
\next}
\def\makeNormal#1#2{\def\doNormalDef{\edef#1}\begingroup%
\aftergroup\doNormalDef\aftergroup{\normalize#2\terminator\aftergroup}%
\endgroup}

\def\warnIfChanged#1#2{%
\ifundef#1
\else\begingroup%
\edef\oldDefinitionOfCounter{#1}\edef\newDefinitionOfCounter{#2}%
\ifx\oldDefinitionOfCounter\newDefinitionOfCounter%
\else%
\linemessage{Warning: definition of \noexpand#1 has changed.}%
\global\AnyCounterChangedtrue\fi\endgroup\fi}

\def\Section#1{\global\advance\secno by1\relax\global\meqno=1%
\global\subsecno=0%
\bigbreak\bigskip
\centerline{\twelvepoint \bf %
\the\secno. #1}%
\par\nobreak\medskip\nobreak}
\def\tagsection#1{%
\warnIfChanged#1{\the\secno}%
\xdef#1{\the\secno}%
\ifWritingAuxFile\immediate\write\auxfile{\noexpand\xdef\noexpand#1{#1}}\fi%
}
\def\section{\Section}
\def\Subsection#1{\global\advance\subsecno by1\relax\medskip %
\leftline{\bf\the\secno.\the\subsecno\ #1}%
\par\nobreak\smallskip\nobreak}
\def\tagsubsection#1{%
\warnIfChanged#1{\the\secno.\the\subsecno}%
\xdef#1{\the\secno.\the\subsecno}%
\ifWritingAuxFile\immediate\write\auxfile{\noexpand\xdef\noexpand#1{#1}}\fi%
}

\def\subsection{\Subsection}

\def\romappno{\uppercase\expandafter{\romannumeral\appno}}
\def\makeNormalizedRomappno{%
\expandafter\makeNormal\expandafter\normalizedromappno%
\expandafter{\romannumeral\appno}%
\edef\normalizedromappno{\uppercase{\normalizedromappno}}}
\def\Appendix#1{\global\advance\appno by1\relax\global\meqno=1\global\secno=0
\bigbreak\bigskip
\centerline{\twelvepoint \bf Appendix %
\romappno. #1}%
\par\nobreak\medskip\nobreak}
\def\tagappendix#1{\makeNormalizedRomappno%
\warnIfChanged#1{\normalizedromappno}%
\xdef#1{\normalizedromappno}%
\ifWritingAuxFile\immediate\write\auxfile{\noexpand\xdef\noexpand#1{#1}}\fi%
}
\def\appendix{\Appendix}

\def\eqn#1{\makeNormalizedRomappno%
\ifnum\secno>0%
  \warnIfChanged#1{\the\secno.\the\meqno}%
  \eqno(\the\secno.\the\meqno)\xdef#1{\the\secno.\the\meqno}%
     \global\advance\meqno by1
\else\ifnum\appno>0%
  \warnIfChanged#1{\normalizedromappno.\the\meqno}%
  \eqno({\rm\romappno}.\the\meqno)%
      \xdef#1{\normalizedromappno.\the\meqno}%
     \global\advance\meqno by1
\else%
  \warnIfChanged#1{\the\meqno}%
  \eqno(\the\meqno)\xdef#1{\the\meqno}%
     \global\advance\meqno by1
\fi\fi%
\eqlabeL#1%
\ifWritingAuxFile\immediate\write\auxfile{\noexpand\xdef\noexpand#1{#1}}\fi%
}
\def\defeqn#1{\makeNormalizedRomappno%
\ifnum\secno>0%
  \warnIfChanged#1{\the\secno.\the\meqno}%
  \xdef#1{\the\secno.\the\meqno}%
     \global\advance\meqno by1
\else\ifnum\appno>0%
  \warnIfChanged#1{\normalizedromappno.\the\meqno}%
  \xdef#1{\normalizedromappno.\the\meqno}%
     \global\advance\meqno by1
\else%
  \warnIfChanged#1{\the\meqno}%
  \xdef#1{\the\meqno}%
     \global\advance\meqno by1
\fi\fi%
\eqlabeL#1%
\ifWritingAuxFile\immediate\write\auxfile{\noexpand\xdef\noexpand#1{#1}}\fi%
}
\def\anoneqn{\makeNormalizedRomappno%
\ifnum\secno>0
  \eqno(\the\secno.\the\meqno)%
     \global\advance\meqno by1
\else\ifnum\appno>0
  \eqno({\rm\normalizedromappno}.\the\meqno)%
     \global\advance\meqno by1
\else
  \eqno(\the\meqno)%
     \global\advance\meqno by1
\fi\fi%
}
\def\mfig#1#2{\global\advance\figno by1%
\relax#1\the\figno%
\warnIfChanged#2{\the\figno}%
\edef#2{\the\figno}%
\reflabeL#2%
\ifWritingAuxFile\immediate\write\auxfile{\noexpand\xdef\noexpand#2{#2}}\fi%
}

\catcode`@=11 

\font\ninerm=cmr9
\font\eightrm=cmr8
\font\sixrm=cmr6

\def\loadtrueseventeenpoint{
 \font\seventeenrm=cmr10 at 17.28truept
 \font\seventeeni=cmmi10 at 17.28truept
 \font\seventeenbf=cmbx10 at 17.28truept
 \font\seventeenit=cmti10 at 17.28truept
 \font\seventeensl=cmsl10 at 17.28truept
 \font\seventeensy=cmsy10 at 17.28truept
}
\def\loadfourteenpoint{
\font\fourteenrm=cmr10 at 14.4pt
\font\fourteeni=cmmi10 at 14.4pt
\font\fourteenit=cmti10 at 14.4pt
\font\fourteensl=cmsl10 at 14.4pt
\font\fourteensy=cmsy10 at 14.4pt
\font\fourteenbf=cmbx10 at 14.4pt
}
\def\loadtruetwelvepoint{
\font\twelverm=cmr10 at 12truept
\font\twelvei=cmmi10 at 12truept
\font\twelveit=cmti10 at 12truept
\font\twelvesl=cmsl10 at 12truept
\font\twelvesy=cmsy10 at 12truept
\font\twelvebf=cmbx10 at 12truept
}

\font\ninei=cmmi9
\font\eighti=cmmi8
\font\sixi=cmmi6
\skewchar\ninei='177 \skewchar\eighti='177 \skewchar\sixi='177

\font\ninesy=cmsy9
\font\eightsy=cmsy8
\font\sixsy=cmsy6
\skewchar\ninesy='60 \skewchar\eightsy='60 \skewchar\sixsy='60

\font\ninebf=cmbx9
\font\eightbf=cmbx8
\font\sixbf=cmbx6

\font\ninett=cmtt9
\font\eighttt=cmtt8

\hyphenchar\tentt=-1 
\hyphenchar\ninett=-1
\hyphenchar\eighttt=-1

\font\ninesl=cmsl9
\font\eightsl=cmsl8

\font\nineit=cmti9
\font\eightit=cmti8


\newskip\ttglue
\def\tenpoint{\def\rm{\fam0\tenrm}%
  \textfont0=\tenrm \scriptfont0=\sevenrm \scriptscriptfont0=\fiverm
  \textfont1=\teni \scriptfont1=\seveni \scriptscriptfont1=\fivei
  \textfont2=\tensy \scriptfont2=\sevensy \scriptscriptfont2=\fivesy
  \textfont3=\tenex \scriptfont3=\tenex \scriptscriptfont3=\tenex
  \def\it{\fam\itfam\tenit}\textfont\itfam=\tenit
  \def\sl{\fam\slfam\tensl}\textfont\slfam=\tensl
  \def\bf{\fam\bffam\tenbf}\textfont\bffam=\tenbf \scriptfont\bffam=\sevenbf
  \scriptscriptfont\bffam=\fivebf
  \normalbaselineskip=12pt
  \let\sc=\eightrm
  \let\big=\tenbig
  \setbox\strutbox=\hbox{\vrule height8.5pt depth3.5pt width\z@}%
  \normalbaselines\rm}

\def\twelvepoint{\def\rm{\fam0\twelverm}%
  \textfont0=\twelverm \scriptfont0=\ninerm \scriptscriptfont0=\sevenrm
  \textfont1=\twelvei \scriptfont1=\ninei \scriptscriptfont1=\seveni
  \textfont2=\twelvesy \scriptfont2=\ninesy \scriptscriptfont2=\sevensy
  \textfont3=\tenex \scriptfont3=\tenex \scriptscriptfont3=\tenex
  \def\it{\fam\itfam\twelveit}\textfont\itfam=\twelveit
  \def\sl{\fam\slfam\twelvesl}\textfont\slfam=\twelvesl
  \def\bf{\fam\bffam\twelvebf}\textfont\bffam=\twelvebf
\scriptfont\bffam=\ninebf
  \scriptscriptfont\bffam=\sevenbf
  \normalbaselineskip=12pt
  \let\sc=\eightrm
  \let\big=\tenbig
  \setbox\strutbox=\hbox{\vrule height8.5pt depth3.5pt width\z@}%
  \normalbaselines\rm}

\def\fourteenpoint{\def\rm{\fam0\fourteenrm}%
  \textfont0=\fourteenrm \scriptfont0=\tenrm \scriptscriptfont0=\sevenrm
  \textfont1=\fourteeni \scriptfont1=\teni \scriptscriptfont1=\seveni
  \textfont2=\fourteensy \scriptfont2=\tensy \scriptscriptfont2=\sevensy
  \textfont3=\tenex \scriptfont3=\tenex \scriptscriptfont3=\tenex
  \def\it{\fam\itfam\fourteenit}\textfont\itfam=\fourteenit
  \def\sl{\fam\slfam\fourteensl}\textfont\slfam=\fourteensl
  \def\bf{\fam\bffam\fourteenbf}\textfont\bffam=\fourteenbf%
  \scriptfont\bffam=\tenbf
  \scriptscriptfont\bffam=\sevenbf
  \normalbaselineskip=17pt
  \let\sc=\elevenrm
  \let\big=\tenbig
  \setbox\strutbox=\hbox{\vrule height8.5pt depth3.5pt width\z@}%
  \normalbaselines\rm}

\def\seventeenpoint{\def\rm{\fam0\seventeenrm}%
  \textfont0=\seventeenrm \scriptfont0=\fourteenrm \scriptscriptfont0=\tenrm
  \textfont1=\seventeeni \scriptfont1=\fourteeni \scriptscriptfont1=\teni
  \textfont2=\seventeensy \scriptfont2=\fourteensy \scriptscriptfont2=\tensy
  \textfont3=\tenex \scriptfont3=\tenex \scriptscriptfont3=\tenex
  \def\it{\fam\itfam\seventeenit}\textfont\itfam=\seventeenit
  \def\sl{\fam\slfam\seventeensl}\textfont\slfam=\seventeensl
  \def\bf{\fam\bffam\seventeenbf}\textfont\bffam=\seventeenbf%
  \scriptfont\bffam=\fourteenbf
  \scriptscriptfont\bffam=\twelvebf
  \normalbaselineskip=21pt
  \let\sc=\fourteenrm
  \let\big=\tenbig
  \setbox\strutbox=\hbox{\vrule height 12pt depth 6pt width\z@}%
  \normalbaselines\rm}

\def\ninepoint{\def\rm{\fam0\ninerm}%
  \textfont0=\ninerm \scriptfont0=\sixrm \scriptscriptfont0=\fiverm
  \textfont1=\ninei \scriptfont1=\sixi \scriptscriptfont1=\fivei
  \textfont2=\ninesy \scriptfont2=\sixsy \scriptscriptfont2=\fivesy
  \textfont3=\tenex \scriptfont3=\tenex \scriptscriptfont3=\tenex
  \def\it{\fam\itfam\nineit}\textfont\itfam=\nineit
  \def\sl{\fam\slfam\ninesl}\textfont\slfam=\ninesl
  \def\bf{\fam\bffam\ninebf}\textfont\bffam=\ninebf \scriptfont\bffam=\sixbf
  \scriptscriptfont\bffam=\fivebf
  \normalbaselineskip=11pt
  \let\sc=\sevenrm
  \let\big=\ninebig
  \setbox\strutbox=\hbox{\vrule height8pt depth3pt width\z@}%
  \normalbaselines\rm}

\def\eightpoint{\def\rm{\fam0\eightrm}%
  \textfont0=\eightrm \scriptfont0=\sixrm \scriptscriptfont0=\fiverm%
  \textfont1=\eighti \scriptfont1=\sixi \scriptscriptfont1=\fivei%
  \textfont2=\eightsy \scriptfont2=\sixsy \scriptscriptfont2=\fivesy%
  \textfont3=\tenex \scriptfont3=\tenex \scriptscriptfont3=\tenex%
  \def\it{\fam\itfam\eightit}\textfont\itfam=\eightit%
  \def\sl{\fam\slfam\eightsl}\textfont\slfam=\eightsl%
  \def\bf{\fam\bffam\eightbf}\textfont\bffam=\eightbf \scriptfont\bffam=\sixbf%
  \scriptscriptfont\bffam=\fivebf%
  \normalbaselineskip=9pt%
  \let\sc=\sixrm%
  \let\big=\eightbig%
  \setbox\strutbox=\hbox{\vrule height7pt depth2pt width\z@}%
  \normalbaselines\rm}

\def\tenbig#1{{\hbox{$\left#1\vbox to8.5pt{}\right.\n@space$}}}
\def\ninebig#1{{\hbox{$\textfont0=\tenrm\textfont2=\tensy
  \left#1\vbox to7.25pt{}\right.\n@space$}}}
\def\eightbig#1{{\hbox{$\textfont0=\ninerm\textfont2=\ninesy
  \left#1\vbox to6.5pt{}\right.\n@space$}}}

\def\footnote#1{\edef\@sf{\spacefactor\the\spacefactor}#1\@sf
      \insert\footins\bgroup\eightpoint
      \interlinepenalty100 \let\par=\endgraf
        \leftskip=\z@skip \rightskip=\z@skip
        \splittopskip=10pt plus 1pt minus 1pt \floatingpenalty=20000
        \smallskip\item{#1}\bgroup\strut\aftergroup\@foot\let\next}
\skip\footins=12pt plus 2pt minus 4pt 
\dimen\footins=30pc 

\newinsert\margin
\dimen\margin=\maxdimen
\def\titlefont{\seventeenpoint}
\loadtruetwelvepoint 
\loadtrueseventeenpoint
\catcode`\@=\active
\catcode`@=12  
\catcode`\"=\active

\def\eatOne#1{}
\def\ifundef#1{\expandafter\ifx%
\csname\expandafter\eatOne\string#1\endcsname\relax}
\def\notTrue{\iffalse}\def\isTrue{\iftrue}
\def\ifdef#1{{\ifundef#1%
\aftergroup\notTrue\else\aftergroup\isTrue\fi}}
\def\use#1{\ifundef#1\linemessage{Warning: \string#1 is undefined.}%
{\tt \string#1}\else#1\fi}


\global\newcount\refno \global\refno=1
\newwrite\rfile
\newlinechar=`\^^J
\def\ref#1#2{\the\refno\nref#1{#2}}
\def\nref#1#2{\xdef#1{\the\refno}%
\ifnum\refno=1\immediate\openout\rfile=\jobname.refs\fi%
\immediate\write\rfile{\noexpand\item{[\noexpand#1]\ }#2.}%
\global\advance\refno by1}
\def\lref#1#2{\the\refno\xdef#1{\the\refno}%
\ifnum\refno=1\immediate\openout\rfile=\jobname.refs\fi%
\immediate\write\rfile{\noexpand\item{[\noexpand#1]\ }#2\semi}%
\global\advance\refno by1}
\def\cref#1{\immediate\write\rfile{#1\semi}}

\def\semi{;\hfil\noexpand\break}

\def\listrefs{\vfill\eject\immediate\closeout\rfile
\centerline{{\bf References}}\bigskip\frenchspacing%
\input \jobname.refs\vfill\eject\nonfrenchspacing}

\def\inputAuxIfPresent#1{\immediate\openin1=#1
\ifeof1\message{No file \auxfileName; I'll create one.
}\else\closein1\relax\input\auxfileName\fi%
}

\newif\ifWritingAuxFile
\newwrite\auxfile
\def\SetUpAuxFile{%
\xdef\auxfileName{\jobname.aux}%
\inputAuxIfPresent{\auxfileName}%
\WritingAuxFiletrue%
\immediate\openout\auxfile=\auxfileName}

\def\L{\left(}\def\R{\right)}

\def\bye{\par\vfill\supereject%
\ifAnyCounterChanged\linemessage{
Some counters have changed.  Re-run tex to fix them up.}\fi%
\end}


\def\Tr{\mathop{\rm Tr}\nolimits}

\def\A#1{{\cal A}_{#1}}

\def\pol{\varepsilon}

\def\c{\,\cdot\,}

\def\L{\left(}\def\R{\right)}

\def\spa#1.#2{\left\langle#1\,#2\right\rangle}
\def\spb#1.#2{\left[#1\,#2\right]}
\def\lor#1.#2{\left(#1\,#2\right)}
\def\sand#1.#2.#3{%
\left\langle\smash{#1}{\vphantom1}^{-}\right|{#2}%
\left|\smash{#3}{\vphantom1}^{-}\right\rangle}
\def\sandp#1.#2.#3{%
\left\langle\smash{#1}{\vphantom1}^{-}\right|{#2}%
\left|\smash{#3}{\vphantom1}^{+}\right\rangle}
\def\sandpp#1.#2.#3{%
\left\langle\smash{#1}{\vphantom1}^{+}\right|{#2}%
\left|\smash{#3}{\vphantom1}^{+}\right\rangle}
\catcode`@=11  
\def\meqalign#1{\,\vcenter{\openup1\jot\m@th
   \ialign{\strut\hfil$\displaystyle{##}$ && $\displaystyle{{}##}$\hfil
             \crcr#1\crcr}}\,}
\catcode`@=12  

\loadfourteenpoint

\leftlabelstrue

%
\overfullrule 0pt
\hfuzz 35 pt
\vbadness=10001
%
%

%

\def\lr{\leftrightarrow}

\def\Slash#1{\slash\hskip -0.17 cm #1}

\def\Split{\mathop{\rm Split}\nolimits}

\def\tree{{\rm tree}}
\def\Gr{{\rm Gr}}
%

\baselineskip 15pt
\overfullrule 0.5pt


\def\ref{\nref}

\ref\Ellis{R.K. Ellis and J.C. Sexton, Nucl.\ Phys.\ B269:445 (1986)}

\ref\ParkeTaylor{S.J. Parke and T.R. Taylor, Phys.\ Rev.\ Lett.\ 56:2459
(1986)}

\ref\RecursiveA{F.A. Berends and W.T. Giele, Nucl.\ Phys.\ B306:759 (1988)}

\ref\FiveGluon{Z. Bern, L. Dixon and D.A. Kosower, Phys.\ Rev. Lett.\
70:2677 (1993)}

\ref\Tasi{
Z. Bern, hep-ph/9304249, in {\it Proceedings of Theoretical
Advanced Study Institute in High Energy Physics (TASI 92)},
eds.\ J. Harvey and J. Polchinski (World Scientific, 1993)}

\ref\SusyFour{Z. Bern, D. Dunbar, L. Dixon and D. Kosower, preprint
hep-ph/9403226, to appear in Nucl.\ Phys.\ B; in preparation}

\ref\AllPlus{Z. Bern, G. Chalmers, L. Dixon and D.A. Kosower,
Phys.\ Rev.\ Lett.\ 72:2134 (1994)}

\ref\Mahlon{G.D.\ Mahlon, Phys.\ Rev.\ D49:2197 (1994);
preprint Fermilab-Pub-93/389-T, hep-ph/9312276}

\ref\StringBased{
Z. Bern and D.A.\ Kosower, Phys.\ Rev.\ Lett.\ 66:1669 (1991);
Nucl.\ Phys.\ B379:451 (1992)\semi {\it Proceedings of the PASCOS-91
Symposium}, eds.\ P. Nath and S. Reucroft (World Scientific, 1992)\semi
Z. Bern, Phys.\ Lett.\ 296B:85 (1992)\semi
Z. Bern and D.C.\ Dunbar,  Nucl.\ Phys.\ B379:562 (1992)\semi
Z. Bern, D.C. Dunbar and T. Shimada, Phys.\ Lett.\ 312B:277 (1993)}

\ref\TreeColor{F.A. Berends and W.T. Giele,
Nucl.\ Phys.\ B294:700 (1987)\semi
M.\ Mangano, S. Parke, and Z.\ Xu, Nucl.\ Phys.\ B298:653 (1988)\semi
Mangano, Nucl.\ Phys.\ B309:461 (1988)}

\ref\Color{Z. Bern and D.A.\ Kosower, Nucl.\ Phys.\ B362:389 (1991)}

\ref\SpinorHelicity{
F.A.\ Berends, R.\ Kleiss, P.\ De Causmaecker, R.\ Gastmans and T.\ T.\ Wu,
        Phys.\ Lett.\ 103B:124 (1981)\semi
P.\ De Causmaeker, R.\ Gastmans,  W.\ Troost and  T.T.\ Wu,
Nucl. Phys. B206:53 (1982)\semi
R.\ Kleiss and W.J.\ Stirling,
   Nucl.\ Phys.\ B262:235 (1985)\semi
   J.F.\ Gunion and Z.\ Kunszt, Phys.\ Lett.\ 161B:333 (1985)\semi
Z. Xu, D.-H.\ Zhang and L. Chang, Nucl.\ Phys.\ B291:392 (1987)}

\ref\ManganoReview{M. Mangano and S.J. Parke, Phys.\ Rep.\ 200:301 (1991)}

\ref\Susy{M.T.\ Grisaru, H.N.\ Pendleton and P.\ van Nieuwenhuizen,
Phys. Rev. {D15}:996 (1977)\semi
M.T. Grisaru and H.N. Pendleton, Nucl.\ Phys.\ B124:81 (1977)}

\ref\DP{L. Dixon, unpublished; G. Mahlon, private communication}


\noindent
hep-ph/9405393  \hfill UCLA/94/TEP/24

\vskip -.8 in
\Title{Constructing QCD Loop Amplitudes Using Collinear Limits%
\footnote{${}^*$}%
{Talk presented at XXII ITEP International Winter School of Physics,
February~22 -- March~2, 1994}%
}

\vskip -.7 cm
\centerline{Gordon Chalmers}
\baselineskip12truept
\centerline{\it Department of Physics}
\centerline{\it University of California, Los Angeles}
\centerline{\it Los Angeles, CA 90024}

\vskip 0.2in\baselineskip13truept

\ifdraftmode
\vskip 5pt
\centerline{{\bf Draft}\hskip 10pt\TimeStamp}
\vskip 5pt
\fi

\centerline{\bf Abstract}
{\ninerm
\narrower We discuss how higher-point QCD amplitudes may be
constructed from lower point ones by imposing the factorization
constraints in the limits as external momenta become collinear. As a
particular example, the all-$n$ gluon one-loop amplitude with maximal
helicity violation is presented.  We also discuss the necessary
collinear behavior of the $n$-gluon amplitudes.}

\vskip .1 in
\baselineskip17pt

\noindent{\bf 1. Introduction.}    The task of finding new
physics in future experiments requires an accurate understanding
of the perturbative QCD background.  In jet physics, for example,
one needs the loop corrections in order to reduce theoretical
uncertainties in the cone angle and the renormalization
scale dependence.  However, the algebraic complexity
of loop calculations in gauge theories prohibits the practical use
of conventional techniques for large numbers of partons.  Even
a four-point one-loop amplitude is formidable with conventional
Feynman diagrammatic techniques [\use\Ellis].  Indirect means
may alternatively be used to find amplitudes,
without any Feynman diagrams.  For example, at tree-level
concise formulae for maximally helicity violating amplitudes
with an arbitrary number of external legs were first
conjectured by Parke and Taylor [\use\ParkeTaylor] and later
proven by Berends and Giele [\use\RecursiveA] using recursion
relations.  The result was found in part by examining the
collinear behavior of tree-level amplitudes, the
universal behavior of which is independent of the number of
external legs.

At one-loop order the QCD gluon amplitudes have a natural
decomposition in terms of supersymmetric contributions plus
a complex scalar loop piece [\use\FiveGluon,\use\Tasi].  The
supersymmetric parts are most easily obtained [\use\SusyFour]
from unitarity through the use of Cutkosky rules.  The
scalar contribution, however, contains polynomials which
cannot be determined in the same manner.  Our goal here is
to show how the universal behavior of the loop level collinear
limits may be used to find scalar pieces of one-loop gluon
amplitudes.

The consistency conditions which one-loop gauge theory
amplitudes satisfy in the limits when the external momenta
become collinear or soft are strong enough to determine
certain amplitudes without using any Feynman diagrams.
In this talk we discuss an example of a one-loop
amplitude which is sufficiently constrained that we can
write down a form for an arbitrary number of external
legs.  The particular all-$n$ result which we present
[\use\AllPlus] is for maximal helicity violation, that is
with all (outgoing) legs of identical helicity, and has since
been confirmed by recursive techniques [\use\Mahlon].  The
construction is based upon extending the known
one-loop four- and five-gluon [\use\FiveGluon] amplitudes which
were obtained using string-based methods [\use\StringBased].

In order to use the collinear limits to fix the polynomial
terms, one needs a proof that the scalar contributions to
the loop have a universal behavior for any number of external
legs as the momenta of two of the legs become collinear.  We
discuss such a proof here.  Since the cuts can be used to
determine the amplitudes completely except for the polynomial
part of the scalar loop piece, we only focus here on the
collinear limits of these scalar loop contributions.

\vskip .1 truein
\noindent{\bf 2. Review.}   We first briefly review some of the
important tools relevant to the content of the present work:
color-ordering the non-abelian Feynman rules, the use of a
spinor helicity basis, and the development of string-improved methods.

Color ordering the Feynman rules amounts to separating a given
amplitude into independent color structures, which must then be
individually gauge invariant.  In the adjoint representation and at
tree level, we may rewrite the structure constants in all the
vertices as $f_{abc} = -i /\sqrt{2} {\rm Tr}\Bigl( [T^a,T^b], T^c
\Bigr)$ and by using trace identities write the full $n$-gluon amplitude as
[\use\TreeColor]:
$$
\A{n}^\tree(\{k_i,\lambda_i,a_i\}) =
g^{n-2} \sum_{\sigma\in S_n/Z_n} \Tr(T^{a_{\sigma(1)}}
\cdots T^{a_{\sigma(n)}})
\ A_n^\tree(k_{\sigma(1)}^{\lambda_{\sigma(1)}},\ldots,
            k_{\sigma(n)}^{\lambda_{\sigma(n)}})\ ,
\eqn\TreeAmplitudeDecomposition
$$
where $k_i$, $\lambda_i$, and $a_i$ are respectively the momentum,
helicity ($\pm$), and color index of the $i$-th external
gluon.  The coupling constant is $g$, and $S_n/Z_n$ is the set of
non-cyclic permutations of $\{1,\ldots, n\}$.  When calculating the
full amplitude we only have to concentrate on finding the simpler
color-ordered partial amplitudes $A_{n}$.  The full contribution
is simply a sum over all inequivalent orderings of
the color traces times ordered amplitudes.

The corresponding color ordering for the adjoint representation
at loop level is slightly more complicated [\use\Color], giving:
$$
{\cal A}_n\L \{k_i,\lambda_i,a_i\}\R =
 g^n \sum_{J} n_J\,\sum_{c=1}^{\lfloor{n/2}\rfloor+1}
      \sum_{\sigma \in S_n/S_{n;c}}
     \Gr_{n;c}\L \sigma \R\,A_{n;c}^{[J]}(\sigma),
\eqn\ColorDecomposition
$$
where ${\lfloor{x}\rfloor}$ is the largest integer less than or
equal to $x$ and $n_J$ is the number of particles of spin $J$.
The leading color-structure factor,
$\Gr_{n;1}(1) = N_c\ \Tr\L T^{a_1}\cdots T^{a_n}\R$,
is just the number of colors, $N_c$, times the tree color
factor.  The subleading color structures ($c>1)$ are given by
$\Gr_{n;c}(1) = \Tr\L T^{a_1}\cdots T^{a_{c-1}}\R\,
\Tr\L T^{a_c}\cdots T^{a_n}\R$,
$S_n$ is the set of all permutations of $n$ objects,
and $S_{n;c}$ is the subset leaving $\Gr_{n;c}$ invariant.
For internal particles in the fundamental ($N_c+\bar{N_c}$)
representation, only the single-trace color structure
($c=1$) would be present, and the corresponding color factor
would be smaller by a factor of $N_c$.  In this talk we
concern ourselves only with the partial amplitudes
$A_{n;1}$; the other partial amplitudes $A_{n;c}$ may be
obtained by appropriate permutations over $A_{n;1}$
[\use\SusyFour].

Another advance in the technology of amplitude calculations is
in the use of a spinor helicity basis [\use\SpinorHelicity].
Roughly speaking, there is a large degree
of redundancy in the terms of Feynman diagrams contributing
to an amplitude.  Much of this redundancy can be removed by
utilizing the invariance of the amplitude under $\pol_i
\rightarrow \pol_i + f(k_i) k_i$, and choosing appropriate
gauges.  In the spinor helicity basis all quantities
are written in terms of Weyl spinors $\vert k^{\pm}
\rangle$, which provide a convenient shorthand for expressing
results.  In this formalism, the polarization
vectors are expressed as $\pol_{\alpha}^{+} (p,q) = \langle q^{-}
\vert \gamma_{\alpha} \vert p^{-} \rangle / \sqrt{2} \langle
q^{-} \vert p^{+} \rangle$ and
$\pol_{\alpha}^{-} (p,q) = \langle q^{+} \vert \gamma_{\alpha}
\vert p^{+} \rangle / \sqrt{2} \langle p^{+}\vert q^{-} \rangle$.
The important point is that the reference
momenta $q$, like $f(k_i)$ above, reflects the gauge invariance
of the amplitude and must drop out of any final result (so
choose it advantageously in a given calculation).  For the
purposes of this presentation we note that
$$
\langle k_i^{-} \vert k_j^{+} \rangle \equiv \langle ij \rangle =
\sqrt{2k_i \c k_j} \exp(i\phi), \quad\quad
\langle k_i^{+} \vert k_j^{-} \rangle \equiv [ij] =
\sqrt{2k_i \c k_j} \exp(-i\phi)
\anoneqn
$$
which vanishes in the limit $k_i \c k_j\rightarrow 0$.

More recently, methods based on limits of certain string
theories have advanced the ability to do one-loop perturbative
QCD calculations [\use\StringBased].  This technique has been
used in the calculation of all five-point gluon helicity
amplitudes, which are necessary as a starting point to generate
the all-$n$ amplitude given in this talk.

\vskip .1 truein
\noindent{\bf 3. Collinear Limits.}   We intend to use the
collinear factorization properties of amplitudes to constrain
the scalar part of the $n$-gluon helicity amplitudes.  First
we discuss these particular kinematical limits of the
amplitudes.  The collinear limits of color-ordered one-loop
QCD amplitudes (with spin-$J$ particle content in the loop) are
expected to have the form:
$$
A_{n;1}^{[J]} \mathop{\longrightarrow}^{a \parallel b}
\sum_{\lambda=\pm}  \biggl(
\Split^{\rm tree}_{-\lambda}(a^{\lambda_a},b^{\lambda_b})\,
A_{n-1;1}^{[J]}(\ldots(a+b)^\lambda\ldots)
+\Split^{[J]}_{-\lambda}(a^{\lambda_a},b^{\lambda_b})\,
A_{n-1}^{\rm tree}(\ldots(a+b)^\lambda\ldots) \biggr)
\eqn\loopsplit
$$
in the limit where the momenta $k_a \rightarrow z P$
and $k_b \rightarrow (1-z) P$ with $P = k_a + k_b$ (where
$b=a+1$).  Here $\lambda$ is the helicity of the intermediate
state with momentum $P$, and the splitting functions
$\Split_{-\lambda}$ describe the divergent infrared
behavior.  This is analogous to the form of tree-level
collinear limits [\use\ManganoReview,\use\ParkeTaylor].  All
known one-loop amplitudes satisfy eq.~(\use\loopsplit).  Due
to the supersymmetry decomposition and techniques based on
unitarity, our main interest is in proving the factorization
for the case of scalars in the loop.  For the example of the
all-plus helicity amplitude, the scalar collinear limits are
sufficient because a SUSY Ward identity relates the gluon and
fermion contribution to the scalar one (discussed further below).

In the following, all of the Feynman diagrams contributing
to the scalar contribution $A^{[0]}_{n;1}$ are systematically
studied with the aim of showing that the functions $\Split^{[0]}$ are
independent of $n$.  The diagrams are first categorized into
several sets depending upon the topology of the two external
collinear legs.  The conclusion is that: (a) $\Split^{\rm tree}$
arises from the diagrams in fig. 1, (b) $\Split^{[0]}$ from
the diagrams in fig. 2, and (c) diagrams without {\it explicit}
$1 / k_1 \c k_2$ poles such as in fig. 3 contribute nothing
to the scalar loop splitting functions.  The forthcoming analysis
is broken up accordingly.

\medskip
\noindent {\it a. Contributions to  $\Split^{\rm tree}$.}  Diagrams
where the adjacent collinear legs are located in an external
tree to the scalar loop potentially contribute to $\Split^{\rm
tree}$.  The exceptional case of this classification, fig. 2b,
contributes to $\Split^{[0]}$.  It is not difficult to see that
the only tree Feynman diagrams which contain poles in $1/{(k_1
+ k_2)^2}$, and hence are leading order in the collinear limit,
are those which have a two-particle external tree as in fig. 1.
The hatched circle represents the
loop amplitude $A_{n-1;1}^{[0]}\bigl( 1+2,\ldots \bigr)$.  Out of this
set we acquire the tree splitting functions, which are given in
[\use\ManganoReview], but repeated here for convenience:
$$
A^{\rm fig.~1a} \longrightarrow^{k_1 \Vert k_2}
\sum_{\lambda =\pm 1}
\Split_{-\lambda}^{\rm tree}(1^{\lambda_1},2^{\lambda_2})
\quad A^{[0]}_{n-1;1}\bigl(\ldots,(1+2)^{\lambda},
\ldots\bigr)
$$
$$
\Split^{\rm tree}_{-} (1^{+},2^{+}) = {1\over \sqrt{z(1-z)} \langle
12 \rangle}, \quad \quad \Split^{\rm tree}_{-} (1^{+},2^{-}) = - {z^2 \over
\sqrt{z(1-z)} [12] },
$$
$$
\Split^{\rm tree}_{-} (1^{-},2^{+}) =
- {(1-z)^2 \over \sqrt{z(1-z)} [12] },
\quad \quad
\Split^{\rm tree}_{-} (1^{-},2^{-}) =
\Split^{\rm tree}_{+} (1^{+},2^{+}) = 0 \, .
\eqn\treesplits
$$

\medskip
\noindent{\it b. Contributions to $\Split^{[0]}$.}   Diagrams
with the collinear legs attached to a loop, but with an explicit
collinear pole give rise to the $\Split^{[0]}$ contribution.
The sum of these diagrams, given in figs. 2a-c, give a
result which does not have any singularity as $k_1$ or
$k_2\rightarrow 0$, does not require renormalization, and
only contains a collinear divergence.  Together they contain
the entire contribution to the loop
splitting functions for internal scalars.  We note that the
finiteness (in $1/(d-4)$) follows by the Ward identity which
relates the infinities of the two- and three-point functions.  We
have from figs.\ 2a-c,
$$
\eqalign{
A^{\rm fig~1}(1,2,\ldots) & \longrightarrow^{k_1 \Vert k_2}
{1\over 16 \pi^2} {1\over 6}
\Big( \pol_1 \c \pol_2 - {\pol_1 \c k_2 \pol_2
\c k_1 \over k_1 \c k_2} \Big)
\Big( {1\over \sqrt{2} k_1 \c k_2} \Big) (k_1 - k_2)^{\mu}
A^{tree}_{n-1 \mu}(1+2,\ldots,n) \cr
& \equiv \sum_{\lambda=\pm} \Split_{-\lambda}^{[0]} (1^{\lambda_1},
2^{\lambda_2}) ~ A^{\rm tree}_{n-1} \big( (1+2)^{\lambda},
\ldots,n \bigr) \, ,
\cr}
\anoneqn
$$
\noindent and after using a helicity basis similar to that
in ref. [\use\RecursiveA]:
$$
\Split_{-}^{[0]}(1^{+},2^{+}) = {1 \over 48 \pi^2} \sqrt{z(1-z)}
{1\over \langle 12 \rangle}, \quad \quad
\Split_{+}^{[0]}(1^{+},2^{+}) = {- 1 \over 48 \pi^2} \sqrt{z(1-z)}
{[12] \over \langle 12 \rangle^2},
$$
$$
\Split_{\pm}^{[0]} (1^{+},2^{-}) = \Split_{\pm}^{[0]}(1^{-},2^{+})=0\, .
\eqn\loopsplits
$$

\noindent{\it c. Subleading as $k_1 \c k_2 \rightarrow 0$. \hskip
.3in}   We now argue that the remaining diagrams
do not have any leading collinear poles.
We first classify the diagrams according to:

\item{(1)} One or both of the collinear legs are attached
via a four-vertex with one of its neighbors (not a collinear
partner) to the scalar loop.  Alternatively, one collinear
leg may be directly connected to a loop where the other
collinear leg is part of a tree sewn onto the loop.

\item{(2)} Both collinear legs are attached to a scalar loop by
three-point vertices and are part of a loop with four or more
legs as depicted in fig.\ 3a.

\item{(3)} The collinear legs are attached by a four-point vertex
to a loop as in fig.\ 3b.

The diagrams in set (1) do not have poles for two
reasons.  First, the scalar product $k_1 \cdot k_2$ is prevented
to appear in the loop integration just by the structure of the
momentum flow.  Second, the external trees attached to these
loops do not have any internal lines that vanish in the
collinear limit. There are no $s_{12}$ channels in any tree
attached to a loop of this type.

All the loop diagrams in sets (2) and (3) are also found to
give subleading contributions in the collinear
limit.  Consider the one-loop diagram
$D_n={\mu}^{\pol}({g/\sqrt2})^n G_n$ with all
external gluon legs connected directly to the scalar
loop by three-point vertices:
$$
G_n = \int {d^{d}l
\over (2\pi)^{d}} \left( \prod^{n}_{j=1} {1\over l^2_j } \right)
\left( \prod^n_{k=1} 2 \pol_k \cdot l_k \right)
\hskip 1.5 cm
{\rm with}~~ l_j = l- q_j,~~q_j\equiv\sum^{j}_{a=2}
k_a~({a~{\rm mod}~n}) \, .
\anoneqn
$$

The leading $1/ k_1 \c k_2$ collinear singularities of the
integrand come from a surface of loop momentum
($l=ak_1 + bk_2$) with thickness of the order $k_1 \c
k_2$.  In this region three propagators simultaneously blow up.
(The two points $l=k_2$ and $l=-k_1$ must be considered separately
since a fourth propagators blows up in the integrand, but the
conclusions are the same.)

We examine the integral by first rewriting the integration
in a manner to extract the contribution from the surface
spanning $k_1$ and $k_2$.  This is
accomplished by breaking the loop momentum
into three components $l=l_{\perp}+ \alpha k_1 +\beta k_2$
($l_{\perp} \cdot k_1 = l_{\perp} \cdot k_2 =0$), with the
measure changing as $d^4l=d^2l_{\perp} (2k_1\cdot k_2)
d\alpha d\beta$ -- valid in Minkowski space.  Since the scalar
one-loop diagrams are infrared finite no dimensional
regulator will be used.  In this manner, with
$l_{\perp}\c k_1 = l_{\perp}\cdot k_2 =0$
$$
G_n={1\over (2\pi)^4} \int d^{2}l_{\perp} d\alpha d\beta
(2k_1\cdot k_2) \left( \prod^{n}_{j=1} {1\over (l_{\perp} +\alpha k_1
+\beta k_2 -q_j)^2 } \right) \left( \prod^n_{k=1} 2\pol_k \cdot
(l_{\perp} +\alpha k_1 +\beta k_2 -q_j) \right) \, .
\eqn\LoopDiagram
$$
The potential collinear divergence in the denominator comes
from the three propagators adjacent to and between the
first and second legs, and arises when $l_{\perp}^2$ becomes small
with respect to the squared momentum flowing through the
three propagators.

In order to extract a collinear singularity from the denominator
of the integral, the component perpendicular to the surface,
$l_{\perp}^2$, has to be restricted to roughly $k_1 \c k_2$
-- the `width' of the surface spanned by $\ell=\alpha k_1 +
\beta k_2$.  In euclidean space we define a cutoff so that
$l_{\perp}^2\leq \Lambda^2\sim k_1 \c k_2$.  To leading order,
we approximate the expression for $G_n$ by ignoring $l_{\perp}$
in all but the propagators $j=n,1,2$ in eq.~(\LoopDiagram), and in
all vertices $l_\perp\c\pol_i$ $($with $i\not= 1,2)$; the corrections
are found in a Taylor expansion in $l_\perp$, but these terms are
suppressed since $l_{\perp}^2/q_j^2 \sim k_1\c k_2/q_j^2$.  In this
way we obtain a `triangle' of three propagators which contributes
the potential collinear behavior.

The contribution to the loop diagram (\LoopDiagram)
from the plane spanned by $k_1$ and $k_2$ in this approximation
is found by integrating over $l_{\perp}$ within a region
$l_{\perp}^2 \leq \Lambda^2$,
$$
\eqalign{
G_n & \approx -{i\over (2\pi)^4} \int
d\alpha d\beta \quad (2k_1\cdot k_2) \left(
\prod^{n}_{j\not=n,1,2}
{1\over (\alpha k_1 +\beta k_2 -q_j)^2 } \right) \left(
\prod^n_{k\not=1,2} 2\pol_k \c (\alpha k_1 +\beta k_2 -q_j)
 \right) \cr
& \hskip 2 cm
\times {\pi \over k_1 \c k_2} \Bigl( \pol_1 \c \pol_2
g(\alpha,\beta, \Lambda'^2)
 -  {\pol_1 \c k_2 \pol_2 \cdot k_1 \over k_1 \c k_2}
 f(\alpha,\beta, \Lambda'^2) \Bigr)  \, .
\cr}
\eqn\ApproxDiagram
$$
The functions $f$ and $g$ are dimensionless quantities and
$\Lambda'^2 \equiv {\Lambda^2/ 2k_1 \c k_2} \sim O(1)$.  By a
dimensional argument or through direct integration, we see that
there are no leading singularities in either $f$ or $g$.

Since there are no leading collinear divergences along
the plane of loop momenta spanned by $k_1$ and $k_2$, we
conclude that these graphs do not contribute to the loop
splitting functions in the limit $k_1 \c k_2 \rightarrow 0$.
The same analysis may be applied to any other graph in this
class and to those in the third set.

A modified analysis for the case in which fermions are in the loop
give the same conclusions as for the scalars.  For the gluons in the
loop, however, it turns out that the diagrams in fig. 3a do contain
collinear singularities; also there are additional complications from
infrared singularities.  A more detailed discussion will be presented
elsewhere.  As previously mentioned, for purposes of constructing
higher-point amplitudes via collinear limits, only the scalar loop
contribution is required.  Hence only the scalar splitting functions
are necessary.

\vskip .1  truein
\noindent{\bf 4. All-Plus Amplitude with Arbitrary Numbers of
Legs.}   The all-plus helicity amplitude $A_{n;1}$
is particularly simple; it is cyclicly symmetric, and no
logarithms or other functions containing branch cuts can appear.
This can be seen by considering the cutting rules:
the cut in a given channel is given by
a phase space integral of the product of the two tree amplitudes
obtained from cutting.  One of these tree amplitudes will vanish
for all assignments of helicities on the cut internal legs since
$A_{n}^{\rm tree} (1^\pm,2^+,3^+,\ldots,n^+)=0$, so that in fact
all cuts vanish. Similar reasoning shows that the all plus
helicity loop amplitude does not contain multi-particle poles;
factorizing the amplitude on a multi-particle pole into lower
point tree and loop amplitudes again yields a tree which vanishes
for either helicity of the leg carrying the multi-particle pole.

Another simplifying feature of the all-plus amplitude is
the equality, up to a sign due to statistics, of the
contributions of internal gluons, complex scalars and Weyl
fermions.  This is a consequence of the supersymmetry Ward
identity [\use\Susy] $A^{\rm susy}(1^\pm,2^+,\ldots,n^+) =
0$ for $N=1$ and $N=2$ theories.  Since the $N=1$ supersymmetry
amplitude has one gluon and one gluino circulating in the
loop, the gluino contribution must be equal and opposite to
that of the gluon in order to yield zero for the total;
similarly, the spectrum of an $N=2$ supersymmetric theory
contains two gluinos and one complex scalar in addition to
the gluon, and the vanishing implies the equality of the
contributions of complex scalars and gluons circulating in
the loop (i.e. $A^{[0]}=A^{[1]}=-A^{[1/2]}$).

The starting point in constructing our $n$-point expression is
the known five-point one-loop helicity amplitude [\use\FiveGluon],
$$
\eqalign{
A_{5;1}(1^+,2^+,3^+,4^+,5^+)\  &=\ {iN_p\over 192\pi^2}\,
  {  s_{12}s_{23} + s_{23}s_{34} + s_{34}s_{45} + s_{45}s_{51} +
     s_{51}s_{12}\ +\ \pol(1,2,3,4)
   \over \spa1.2 \spa2.3 \spa3.4 \spa4.5 \spa5.1 }\ ,  \cr }
\eqn\FivePtAmpl
$$
where
$\pol(1,2,3,4) = 4i\varepsilon_{\mu\nu\rho\sigma}
        k_1^\mu k_2^\nu k_3^\rho k_4^\sigma
    \ =\ \spb{1}.{2}\spa{2}.{3}\spb{3}.{4}\spa{4}.{1}
       - \spa{1}.{2}\spb{2}.{3}\spa{3}.{4}\spb{4}.{1}
$, and $N_p$ is the number of color-weighted bosonic states
minus fermionic states circulating in the loop;
for QCD with $n_f$ quarks, $N_p = 2(1-n_f/N)$ with $N=3$.

Using eq.~(\use\loopsplit) and $A^{\rm tree}_n (1^\pm,
2^+, \cdots, n^+) = 0$, we can construct higher point
amplitudes by writing down general forms with only
two particle-poles, and requiring that they have the
correct collinear limits.  To start the procedure one
assumes that the denominator for the six-point amplitude
is $\spa1.2 \spa2.3 \cdots \spa6.1$.  The numerator is a
polynomial of the correct dimensions with coefficients
fixed by the collinear limit constraints to the five-point
amplitude (\use\FivePtAmpl).  Continuing in this way we can
generalize to the all-$n$ result [\use\AllPlus]
$$
\eqalign{
 A_{n;1}(1^+,2^+,\ldots,n^+)\ =\ -{i N_p \over 192\pi^2}\,
\sum_{1\leq i_1 < i_2 < i_3 < i_4 \leq n}
{ {\rm tr} [(1+\gamma_5) \Slash{k}_{i_1} \Slash{k}_{i_2}
\Slash{k}_{i_3} \Slash{k}_{i_4}]
\over \spa1.2 \spa2.3 \cdots \spa{n}.1 }\ ,
}\eqn\allnplus
$$
a form which has been confirmed in [\use\Mahlon].  It is
straightforward to verify that the above all-plus amplitude
satisfies the collinear limits given in eq.~(\use\loopsplit).

Additionally, the collinear limits of loop amplitudes
leads one to suspect that many amplitudes in massless QED and mixed
QCD/QED must vanish, as found for the all-plus amplitude
in QED by Mahlon [\use\Mahlon].  First,
charge conjugation invariance implies that photon amplitudes
with an odd number of legs vanish.  This also implies that
the amplitude with three photons and two gluons
$A_{5;1}(\gamma_1, \gamma_2, \gamma_3, g_4, g_5) =0$, since
this amplitude is proportional to the corresponding photon
amplitude: the two gluons have to be in a color singlet.  Using
the collinear behavior~(\use\loopsplit) leads one to suspect
that the six-point all-plus helicity amplitudes (lacking cuts)
with three photons and three gluons may vanish, and
continuing recursively in this way, that perhaps all amplitudes
with three photons and additional gluons vanish.  However, it is
possible to construct functions containing logs which have
non-singular behavior in all collinear limits [\use\SusyFour];
these non-singular functions are `missed' in the collinear
bootstrap and pose a potential problem to acquiring log and dilog
terms through the collinear approach.  For our case, the
amplitudes are free of cuts so that these type of functions
cannot appear, although one might still worry about cut free
functions with no singularities in any channel.
Nevertheless, we can directly verify that the all-plus amplitudes
with three or more photons vanish.

Amplitudes with $r$ external photons and $(n-r)$ gluons have
a color decomposition similar to that of the pure-gluon
amplitudes, except that charge matrices are set to unity for
the photon legs. The coefficients of these color factors,
$A_{n;1}^{r\gamma}$, are given by appropriate cyclic sums
over the pure-gluon partial amplitudes. One can
write down simple forms for the all-plus partial amplitude
with one or two external photons, and any number of gluons
[\use\AllPlus].  By explicitly performing the sum over
permutations, we find that for three or more external photons
the amplitude vanishes,
$$
A^{\rm loop}_{n>4}(\gamma_1^+, \gamma_2^+, \gamma_3^+, g_4^+,
\ldots, g_n^+)
 =0.
\anoneqn
$$
Since amplitudes with even more photon legs are obtained by
further sums over permutations of legs, the all-plus helicity
amplitudes with three or more photon legs vanish (for $n>4$) in
agreement with the expectation from the collinear limits.  The
same result holds when one of the helicities is reversed
[\use\DP].

\noindent{\bf 5. Conclusions.}   The combined usage of
collinear limits and unitarity provides powerful methods for
further calculations of gauge theory amplitudes.  On the one
hand, the Cutkosky rules fix the logarithmic
parts to amplitudes and miss the contributions of rational
functions (those which have no cuts).  The QCD $n$-gluon
amplitudes can be decomposed into supersymmetric plus scalar
loop contributions; indeed, it has been proved that the cuts
uniquely determine the supersymmetric parts of the gluon one-loop
amplitudes [\use\SusyFour].  The scalar contributions contain
polynomial terms which are not uniquely fixed by the cuts, and
the collinear limit approach is effective in constraining
these contributions.  In this way, these two approaches
complement each other.

Collinear limits also provide a strong check on amplitudes
obtained by other means, such as string-improved [\use\FiveGluon],
recursive [\use\Mahlon], or unitarity techniques [\use\SusyFour].
We expect that the approach based on the collinear limits will be
a useful tool for finding further one-loop amplitudes (e.g.
$n=6,7, \ldots$) for general helicity configurations.

The author would like to thank Zvi Bern for guidance and
collaboration on this work.  Also, the author thanks L.
Dixon and D. Kosower for collaboration, and D. Dunbar for
helpful comments.

\listrefs

\vfill\break

\centerline{\bf Figure Captions.}

\vskip .5 cm

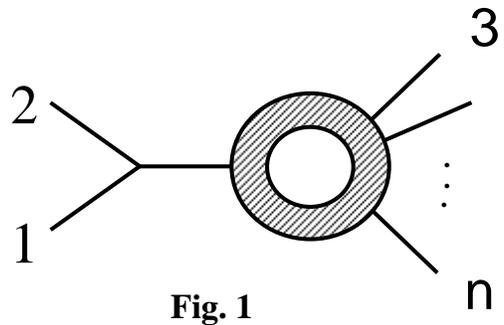
\item{\bf Fig.\ 1:} Diagrams that contribute to the tree splitting functions.

\item{\bf Fig.\ 2:} Diagrams that contribute to the loop splitting functions.

\item{\bf Fig.\ 3:} Two of the remaining diagram types which have no
collinear poles for scalars in the loop.

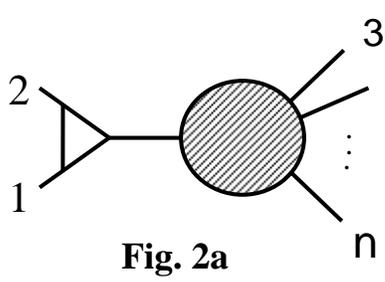
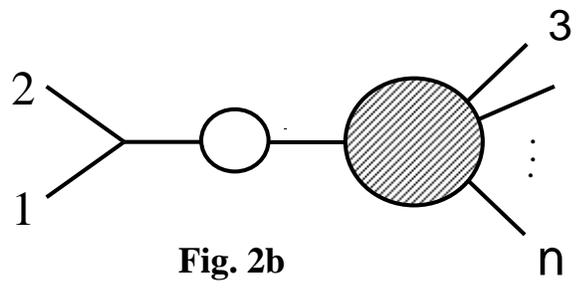
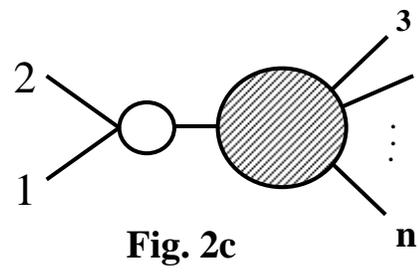
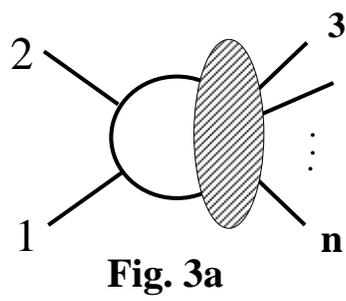
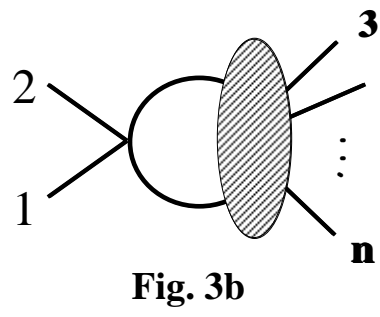
\bye